\documentclass[aps,twocolumn,superscriptaddress]{revtex4}

\usepackage[version=3]{mhchem} 
\usepackage{graphicx}
\usepackage{epstopdf}
\usepackage{color}
\usepackage{dcolumn}
\usepackage{bm}

\begin{document}

\title{Quasi-two-dimensional thermoelectricity in SnSe}

\author{V. Tayari}
\affiliation{Department of Electrical and Computer Engineering, McGill University, Montr\'eal, Qu\'ebec, H3A 2A7, Canada}

\author{B.V. Senkovskiy}
\affiliation{II. Physikalisches Institut, Universit\"at zu K\"oln, Z\"ulpicher Strasse 77, 50937 K\"oln, Germany}

\author{D. Rybkovskiy}
\affiliation{A. M. Prokhorov General Physics Institute, RAS, 38 Vavilov street, 119991, Moscow, Russia}

\author{N. Ehlen}
\affiliation{II. Physikalisches Institut, Universit\"at zu K\"oln, Z\"ulpicher Strasse 77, 50937 K\"oln, Germany}

\author{A. Fedorov}
\affiliation{II. Physikalisches Institut, Universit\"at zu K\"oln, Z\"ulpicher Strasse 77, 50937 K\"oln, Germany}
\affiliation{IFW Dresden, P.O. Box 270116, D-01171 Dresden,
Germany}
\affiliation{St. Petersburg State University, 198504 St. Petersburg, Russia}

\author{C.-Y. Chen}
\affiliation{ANTARES Beamline, Synchrotron SOLEIL \& Universite Paris-Saclay, L' Orme des Merisiers, Saint Aubin-BP 48, 91192 Gif sur Yvette Cedex, France}

\author{J. Avila}
\affiliation{ANTARES Beamline, Synchrotron SOLEIL \& Universite Paris-Saclay, L' Orme des Merisiers, Saint Aubin-BP 48, 91192 Gif sur Yvette Cedex, France}

\author{M. Asensio}
\affiliation{ANTARES Beamline, Synchrotron SOLEIL \& Universite Paris-Saclay, L' Orme des Merisiers, Saint Aubin-BP 48, 91192 Gif sur Yvette Cedex, France}

\author{A. Perucchi}
\affiliation{Elettra-Sincrotrone Trieste S.C.p.A., Area Science Park, I-34012 Trieste, Italy}

\author{P. di Pietro}
\affiliation{Elettra-Sincrotrone Trieste S.C.p.A., Area Science Park, I-34012 Trieste, Italy}

\author{L. Yashina}
\affiliation{Department of Chemistry, Moscow State University, Leninskiye Gory 1/3, 119992, Moscow, Russia}

\author{I. Fakih}
\affiliation{Department of Electrical and Computer Engineering, McGill University, Montr\'eal, Qu\'ebec, H3A 2A7, Canada}

\author{N. Hemsworth}
\affiliation{Department of Electrical and Computer Engineering, McGill University, Montr\'eal, Qu\'ebec, H3A 2A7, Canada}

\author{M. Petrescu}
\affiliation{Department of Physics, McGill University, Montr\'eal, Qu\'ebec, H3A 2A7, Canada}

\author{G. Gervais}
\affiliation{Department of Physics, McGill University, Montr\'eal, Qu\'ebec, H3A 2A7, Canada}

\author{A. Gr\"{u}neis}
\email{grueneis@ph2.uni-koeln.de}
\affiliation{II. Physikalisches Institut, Universit\"at zu K\"oln, Z\"ulpicher Strasse 77, 50937 K\"oln, Germany}

\author{T. Szkopek}
\email{thomas.szkopek@mcgill.ca}
\affiliation{Department of Electrical and Computer Engineering, McGill University, Montr\'eal, Qu\'ebec, H3A 2A7, Canada}
\date{\today}

\begin{abstract}
Stannous selenide is a layered semiconductor that is a polar analogue of black phosphorus, and of great interest as a thermoelectric material. Unusually, hole doped SnSe supports a large Seebeck coefficient at high conductivity, which has not been explained to date. Angle resolved photo-emission spectroscopy, optical reflection spectroscopy and magnetotransport measurements reveal a multiple-valley valence band structure and a quasi two-dimensional dispersion, realizing a Hicks-Dresselhaus thermoelectric contributing to the high Seebeck coefficient at high carrier density. We further demonstrate that the hole accumulation layer in exfoliated SnSe transistors exhibits a field effect mobility of up to $250~\mathrm{cm^2/Vs}$ at $T=1.3~\mathrm{K}$. SnSe is thus found to be a high quality, quasi two-dimensional semiconductor ideal for thermoelectric applications.
\end{abstract}

\maketitle

Layered materials display a remarkable variety of physical properties, and have attracted considerable attention for decades \cite{PALee, VGrasso}. Among the less well studied layered materials is stannous selenide (SnSe), a group IV monochalcogenide whose thermodynamically most stable form at room temperature is an orthorhombic crystal consisting of highly puckered, honeycomb layers with a $d=0.58~\mathrm{nm}$ spacing \cite{Wiedemeier78}. SnSe is a polar analogue of the layered, elemental semiconductor black phosphorus \cite{Morita}, which has been studied in its exfoliated form \cite{Li_NN2014, Liu_ACN2014, Xia_NC2014, Gomez}. Bulk SnSe has attracted much attention recently for its low thermal conductivity $\kappa$\cite{Zhao14, Li2015}, high Seebeck coefficient $S$ \cite{Zhao14} and high electrical conductivity $\sigma$ in hole-doped \cite{Zhao16} and electron-doped \cite{Duong} material. This combination of physical properties leads to a figure of merit $ZT = S^2 \sigma T / \kappa > 2$ for p-type SnSe\cite{Zhao14} and n-type SnSe \cite{Duong}, which is promising for thermoelectric energy conversion applications. A good figure of merit is fundamentally difficult to achieve in a semiconductor because electrical conductivity $\sigma \propto p$ favours a high carrier concentration $p$, while the Seebeck coefficient $S \propto p^{-2/3}$ in a single bulk band favours a low carrier concentration \cite{Snyder}.

The record $ZT$ figure of merit of p-type SnSe has been explained as arising from a valence band structure with multiple low-lying valleys, on the basis of band structure calculations \cite{Zhao16, Cuong_15} that have not been verified by experiment. The role of dimensionality, long recognized as a critical factor in determining density of states and thus Seebeck coefficient \cite{hicks}, has thus far not been considered in the layered material SnSe. Despite its importance for thermoelectric applications, there is paucity of experimental studies of the electronic properties of SnSe. An indirect bandgap of $E_g=0.86-0.94~\mathrm{eV}$ has been measured by optical absorption and reflection experiments in bulk SnSe \cite{maier, yu81, Bhatt_89, Zhao16}. Nominally undoped, bulk SnSe exhibits hole conduction with a room temperature Hall mobility in the range of $150-250~\mathrm{cm^2/Vs}$ \cite{Asanabe, maier, yu81, Zhao16}, reaching $7000~\mathrm{cm^2/Vs}$ at a temperature of 77~K \cite{maier}. Ultra-thin films of poly-crystalline SnSe have been prepared by liquid phase deposition with field effect mobilities of $10~\mathrm{cm^2/Vs}$ \cite{Mitzi}. Vapour phase deposition has been used to synthesize ultra-thin SnSe crystals with field effect mobilities of $1.5-10~\mathrm{cm^2/Vs}$ \cite{Zhao_NR16, Cho17, Pei_16}.

Here, we report a combined spectroscopy and electronic transport study of SnSe single crystals. We performed angle resolved photoelectron spectroscopy (ARPES) measurements of the electronic band structure in the full three-dimensional Brillouin zone, identifying two valleys near band edge, similar to that recently reported \cite{Wang17, Lu17}. One valley is found to be quasi two-dimensional (quasi-2D) with negligible out-of-plane dispersion, and the other valley has an out-of-plane effective mass $0.8 m_0$. Optical reflectance spectroscopy shows a linearly increasing optical conductivity indicative of an indirect bandgap of $0.8~\mathrm{eV}$ and a quasi-2D dispersion in agreement with ARPES. Magnetotransport measurements show excellent agreement with a semi-classical two-carrier model, where the Hall mobility of the high mobility valley reaches $2200~\mathrm{cm^2/Vs}$. The hole accumulation layer in exfoliated SnSe crystals shows modulation with a field effect mobility reaching $250~\mathrm{cm^2/Vs}$ at $T=1.3~\mathrm{K}$. Our experimental study reveals the combined role of valley degeneracy and quasi-2D dispersion in the thermoelectric response of SnSe, and conclusively demonstrates the suitability of SnSe for field effect devices.\\

\begin{figure*}
    \includegraphics [width=0.95\textwidth]{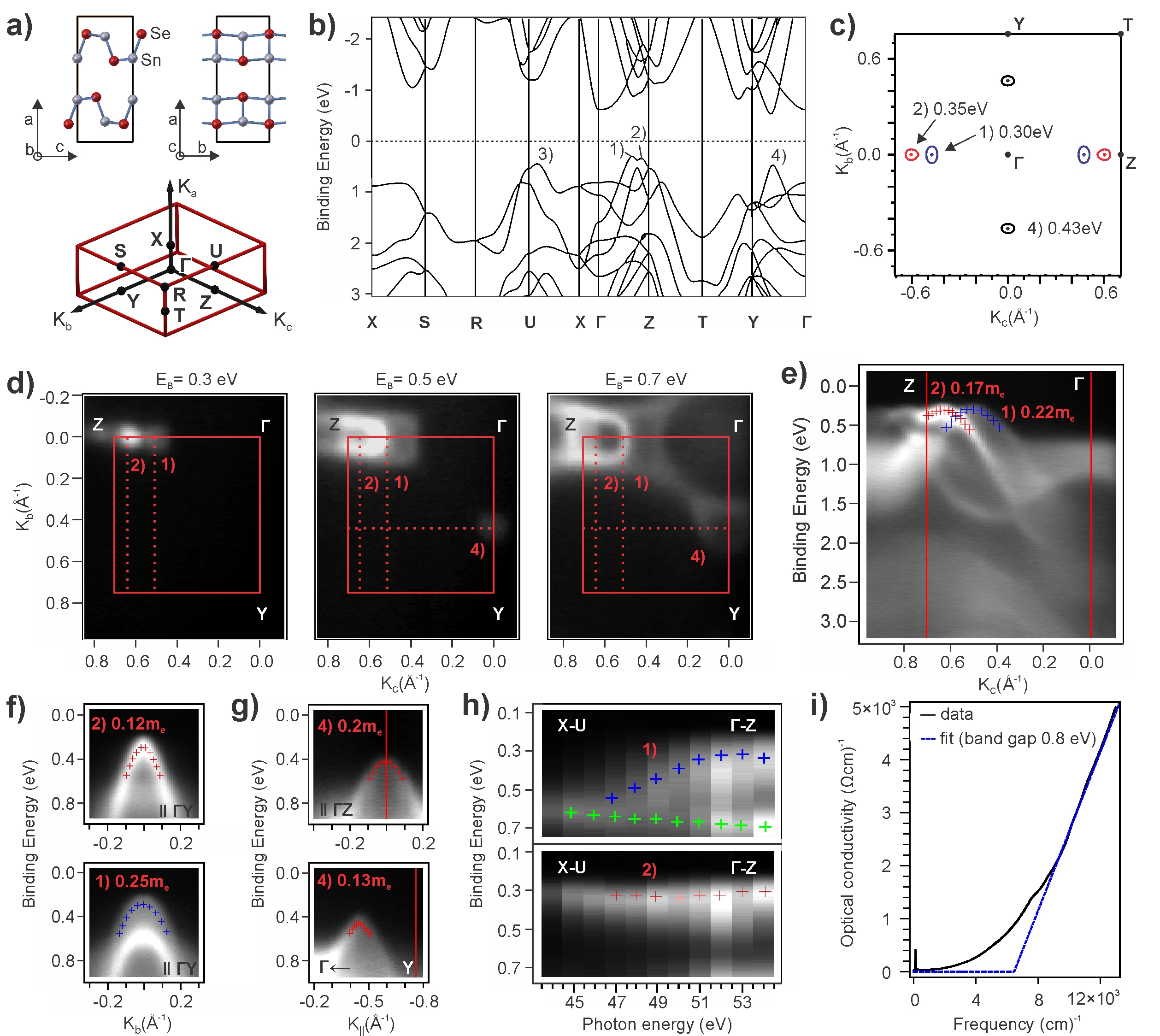}
    \caption{\label{} \textbf{SnSe crystal and band structure.} \textbf{a)} Crystal structure and Brillouin zone (BZ). \textbf{b)} DFT band structure along high symmetry lines, with hole pockets 1) - 4) identified. \textbf{c)} DFT calculated position and shape of the hole pockets in the $\Gamma YZ$ plane of the BZ, with valleys identified. An isoenergy line 0.03 eV below the local valence band maximum is shown for each valley. The energy of each hole valley band maxima is identified. \textbf{d)} ARPES isoenergy cuts at $E=0.3$~eV,$0.5$~eV and $0.7$~eV showing the location of the valleys 1), 2) and 4) in the BZ ($h\nu=74~eV$). \textbf{e)} Momentum cuts along $\Gamma Z$ of valleys 1) and 2). The ARPES intensity maxima and the effective mass are also shown. \textbf{f)} Momentum cuts along $\Gamma Y$ of the pockets 1) and 2) along with the ARPES intensity maxima and the effective masses. \textbf{g)} Momentum cuts parallel to $\Gamma Z$ (upper panel) and along $\Gamma Y$ (lower panel) of pocket 4). \textbf{h)} Out-of-plane ($K_a$) dependence of the ARPES intensities of valleys 1) and 2). The crosses indicate the VBM and allow for extraction of the effective masses along $K_a$ (see text). This momentum cut also shows that valley 3) is the extension of valley 2) along $XU$. \textbf{i)} SnSe optical conductivity spectrum measured at room temperature with unpolarized light at normal incidence. The model fit (blue line) is that for a 2D indirect gap semiconductor.}
\end{figure*}

\textbf{Angle Resolved Photoemission Spectroscopy}\\

The SnSe unit cell, showing a puckered honeycomb layer structure, and the first Brillouin zone (BZ) with high symmetry points are shown in Fig.1(a). The nature of valence band structure in SnSe is critical to understanding the electronic and thermoelectric properties of hole doped SnSe. The band structure predicted by density functional theory (DFT) along the high-symmetry lines of the SnSe BZ is depicted Fig.1(b), with calculational details described in the Appendices below. The Fermi energy was set to 0.30 eV above the valence band maximum (VBM) in line with ARPES measurements of weakly hole doped SnSe described below. An indirect gap is observed, and we identify four hole valleys, 1)-4), in Fig.1(b).

The location and shape of the valence valleys predicted by DFT are visualized in Fig. 1(c) with isoenergy contours 30~meV below each local VBM for valleys 1), 2) and 4) in the $\Gamma YZ$ plane. Valley 3) lies in the $UXS$ plane, and is derived from the out-of-plane dispersion of valley 2) with wavevector in the $K_a$ direction, as seen in Supporting Information Fig.S2 \cite{supp}. In the absence of interlayer coupling, there would be no out-of-plane dispersion, and the band structure around valley 3) would be identical to that around valley 2). As we will show later, the experimental $K_a$ dispersion of valley 2) is indeed negligible and DFT overestimates the $K_a$ dispersion.

According to our ab-initio calculations, valley 1) has a VBM of $E_F-E_1=0.30~\mathrm{eV}$, and valley 2) has a VBM of $E_F-E_2=0.35~\mathrm{eV}$. Valley 4) is lower in energy and is located along $\Gamma Y$ with a VBM of $E_F-E_3=0.43~\mathrm{eV}$. As seen in Fig.1(c), the isoenergy contours are elliptical, indicative of effective mass anisotropy (see Supporting Information \cite{supp} for the effective masses). Subtle changes in the energy and shape of the valleys, below the accuracy of ab-initio calculations, can strongly affect transport and thermoelectric properties, high-lighting the importance of experimental investigation of the band structure using high-resolution ARPES.

We accumulated ARPES intensity maps along the high symmetry crystallographic directions. Fig.1(d) depicts isoenergy contours of the ARPES intensity at three different binding energies (0.3~eV, 0.5~eV and 0.7~eV). The appearance of valleys 1), 2) and 4) is clearly seen, in qualitative agreement with DFT. Valleys 1), 2) and 4) are off zone centre, and are thus doubly degenerate in the language of semiconductors. Cuts along lines of high-symmetry allows extraction of the VBM energy and effective masses for each valley.  Fig.1(e) shows the valley 1) and 2) dispersion along $\Gamma Z$, where it can be seen that the VBM of both are degenerate. The VBM of valley 1) is 0.3~eV below the Fermi level, indicative of weak hole doping in consideration of the 0.8~eV bandgap of SnSe determined by optical spectroscopy below. The anisotropy of the effective masses of valleys 1) and 2) is evident in comparing the momentum cuts parallel to $\Gamma Y$ shown in Fig.1(f) with that along $\Gamma Z$ in Fig.1(e). The dispersion of low-energy valley 4) is shown in Fig. 1(g) parallel to the $\Gamma Z$ and along the $\Gamma Y$ directions. 

The out-of-plane ($K_a$) dispersion of valleys 1) and 2) plays an important role in determining density of states, and hence thermoelectric properties. The oscillations of the position of valley 1) with photon energy allow us to assign the out-of-plane wave vector $K_a$ taking into account the out of plane lattice constant $a=11.49~\AA$ and an inner potential of 12~eV (see Supporting Information \cite{supp}). Fig.1(h) shows the photon energy dependence of the ARPES intensity at valleys 1) and 2). The $\Gamma Z$ line is identified by the maximum splitting of the two highest valence bands at valley 1) and the VBM being at the lowest binding energy. It can be seen from the photon energy dependence of the ARPES intensity maxima that valley 1) has a $K_a$ bandwidth of $\sim$0.3~eV while valley 2) has a negligible dispersion along $K_a$ within an experimental accuracy of about 20~meV. These results indicate that the electronic structure of SnSe is quasi-2D, wherein dispersion out of plane (along $K_a$) is significantly weaker than the in-plane dispersion. Importantly, the $K_a$ dispersion of valley 1) allows us to assign an out-of-plane effective mass at $\Gamma Z$ of about 0.8$m_0$. Notably, these findings are in qualitative agreement with ab-initio calculations which predict 0.55$m_0$ for the out-of-plane effective mass of valley 1) in the $\Gamma YZ$ plane and a 2.5 times higher mass for valley 2) (see Supporting Information \cite{supp}). The finding that valley 2) has a significantly heavier out-of-plane effective mass than valley 1) implies that the density of states and carrier concentration in valley 2) is larger than that of valley 1). We apply this reasoning to the assignment of hole mobility and carrier concentration from magnetotransport measurements discussed below.

Comparing ab-initio predictions with ARPES measurements, it is clear that despite the good agreement at a coarse level, there are important features that are not fully captured by ab-initio calculation. The VBM of valleys 1) and 2) are experimentally found to be equal to within $\Delta < 20~\mathrm{meV}$, while DFT predicts a $\Delta = 50~\mathrm{meV}$ splitting. Similarly, the dispersion along $K_a$ is experimentally found to be flatter than the DFT results. Quantitative modelling of SnSe thermoelectric performance is sensitive to these features. \\

\textbf{Optical Reflectance Spectroscopy}\\

Dimensionality plays an important role in the dependence of optical absorption on photon energy, and the quasi-2D nature of SnSe can also be seen in optical absorption experiments. Fig.1(i) shows the optical conductivity of SnSe inferred from optical reflectivity measurements at nearly normal incidence to the basal plane (alignment with the $a$-axis) of a single crystal of SnSe. The low energy far infrared (IR) region depicts the IR active phonons (not shown) in agreement with previous literature~\cite{cardona77-snse}. At higher energies a linear function can be fit to the conductivity, intersecting the energy axis. The optical conductivity for a 2D indirect band gap semiconductor is proportional to $\hbar\omega + \hbar\omega_0 - E_{gap}$ for $\hbar\omega > E_{gap}-\hbar\omega_0$, where $\hbar\omega$ is the photon energy, $\hbar\omega_0$ the phonon energy absorbed in the indirect optical transition, and $E_{gap}$ the indirect energy gap \cite{PALee}. The absorbed phonon energies are at most 20~meV in SnSe \cite{aamir17-phonons,Li2015}, which is several percent of the previously reported band gap \cite{elkorashy86-snse,martin90-snse} and can be neglected. Thus, a linear fit of the measured optical conductivity to $(\hbar\omega - E_{gap})$ can be used to extract an indirect optical gap of $E_{gap}= 0.80~\mathrm{eV}$, in good agreement with previous works\cite{elkorashy86-snse,martin90-snse}. The optical conductivity below the gap can be attributed to band impurity transitions. \\

\begin{figure}
    \includegraphics [width=3.5in]{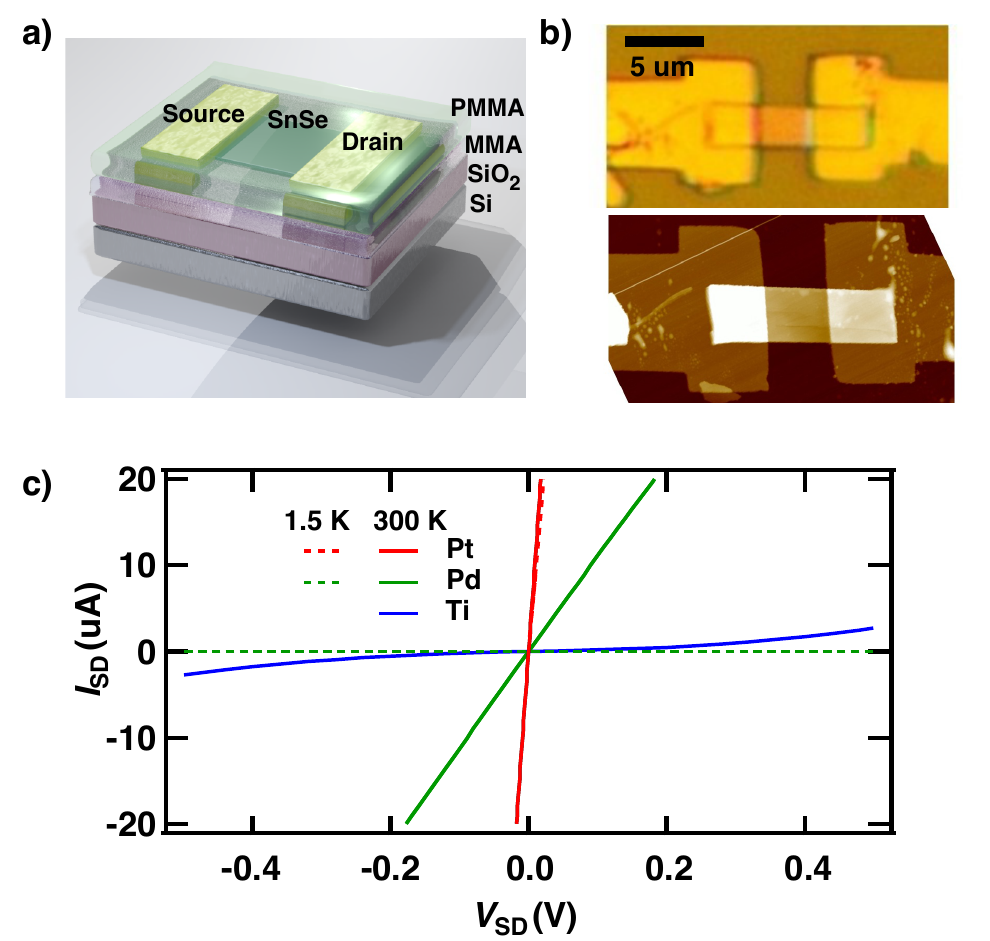}
    \caption{\label{} \textbf{Zero field resistivity.} \textbf{a)} Schematic of exfoliated SnSe FET with Si back-gate. \textbf{b)} Optical micrograph and AFM image of a representative exfoliated SnSe FET with a thickness of 180$\pm$16~nm. \textbf{c)} A comparison of current-voltage characteristics of exfoliated SnSe with Ti, Pd, and Pt contact electrodes, where Pt gives Ohmic conduction at $T$ = 1.5~K and above. }
\end{figure}

\textbf{Device Fabrication}\\

The charge transport properties of exfoliated SnSe flakes and bulk SnSe crystals were investigated in detail. Single crystal SnSe was exfoliated (see Appendix A) on to a degenerately doped Si substrate with 300~nm of SiO$_2$ for back-gating, and electrodes were formed with standard lithography techniques as shown schematically in Fig. 2 a), with a representative device shown in Fig. 2 b). Atomic force microscopy was used to determine exfoliated flake thickness, and devices were found to be in the $\sim100~\mathrm{nm}-400~\mathrm{nm}$ thickness range where quantum confinement effects can be neglected. Polymer encapsulation was used to mitigate against oxidation. The in-plane charge transport of bulk samples cleaved along the basal plane were also investigated, with a typical thickness of $\sim110~\mathrm{\mu m}-170~\mathrm{\mu m}$.

A variety of metals were tested for Ohmic contact formation. The current $I_{SD}$ versus applied voltage $V_{SD}$ of exfoliated SnSe flakes is shown in Fig. 2 c). Titanium contacts resulted in Schottky barrier behaviour at 300~K. Palladium contacts resulted in linear behaviour at 300~K and Schottky behaviour at 1.5~K. In contrast, platinum resulted in linear behaviour at 300~K and at 1.5~K. ARPES, field effect and Hall-effect measurements (described below) all indicate weak, unintentional p-type doping of the SnSe crystals used in our experiments. Although the work function of SnSe remains unknown, the improvement in Ohmic contact formation can nonetheless be understood from the increase in work function for the metal sequence where $\phi = 4.33, 5.22, 5.64~\mathrm{eV}$ for Ti, Pd, and Pt, respectively \cite{CRC}. \\

\begin{figure}
    \includegraphics [width=3.5in]{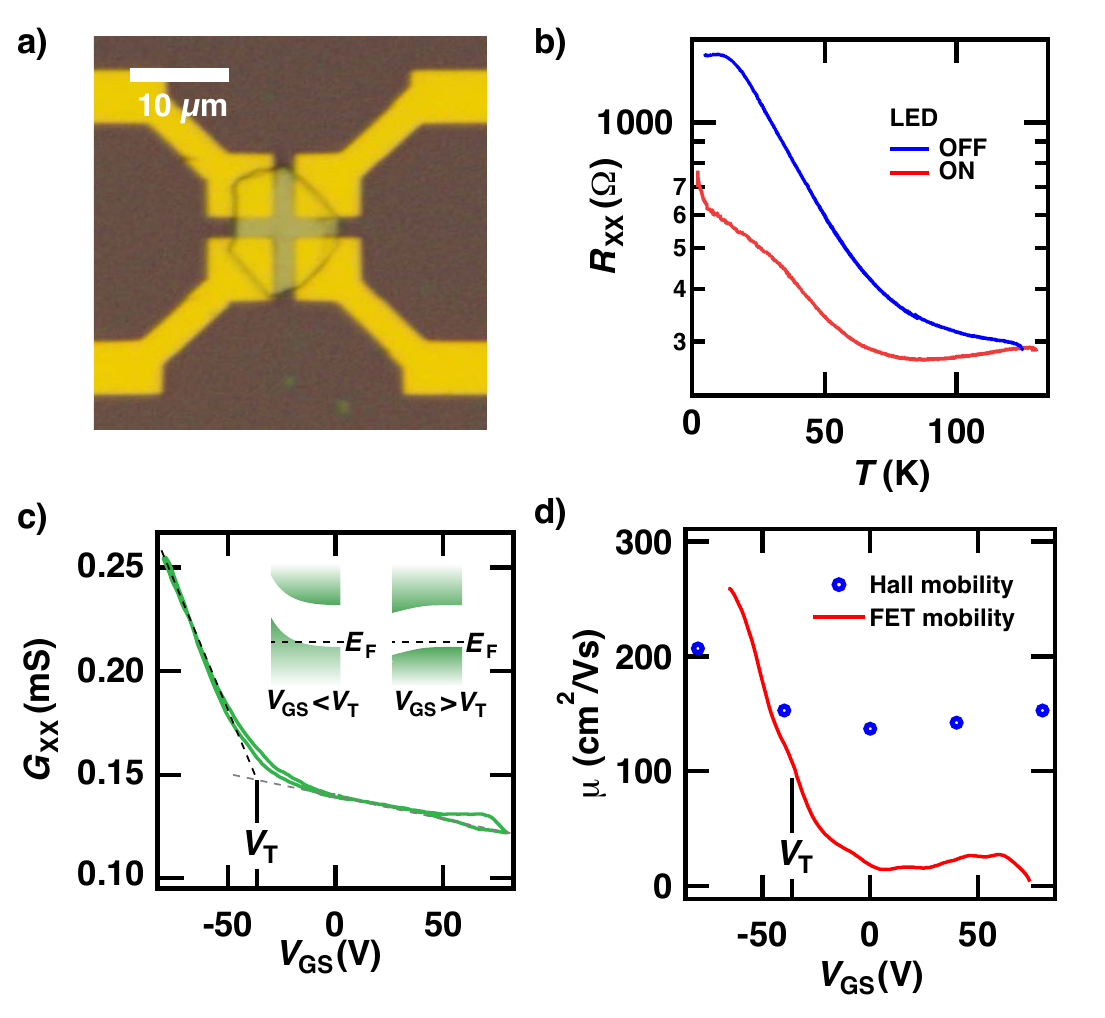}
    \caption{\label{} \textbf{Field Effect.} \textbf{a)} Optical micrograph of an exfoliated SnSe flake with a thickness of $320\pm30~\mathrm{nm}$. \textbf{b)} The resistivity $R_{xx}$ versus temperature $T$ during cool down with and without illumination from a red ( $\lambda = 630~\mathrm{nm}$ ) light emitting diode. \textbf{c)} The sheet conductance $G_{xx}$ versus gate voltage $V_{GS}$ at $T=1.3~\mathrm{K}$. A hole accumulation layer threshold is observed at $V_{GS}=-37~\mathrm{V}$. \textbf{d)} The FET mobility $\mu_{FET}$ and the Hall mobility $\mu_{H}$ versus gate voltage $V_{GS}$ at $T=1.3~\mathrm{K}$.  }
\end{figure}

\textbf{Field Effect}\\

The electronic quality of exfoliated SnSe was confirmed by field effect measurements. A flake in a back-gated van der Pauw geometry, shown in Fig. 3 a), was investigated in depth. The longitudinal resistance $R_{xx}$ versus temperature $T$ is shown in Fig. 3 b), increasing modestly with decreasing temperature. No freeze-out is observed, with conductivity persisting down to a temperature $T=0.3~\mathrm{K}$. Illumination with a red light emitting diode ( $\lambda = 630~\mathrm{nm}$ ) during device cooling resulted in a persistent photoconductivity, similar to that observed at low temperature in other semiconductors \cite{Lang}. We attribute the persistent increase in p-type conductivity to photo-electron trapping, which increases the fixed negative charge density that supports the mobile hole density. The microscopic nature of the electron trap states remains to be identified.

The longitudinal sheet conductance $G_{xx}$, corrected for van der Pauw geometry, is shown in Fig. 3c) versus gate voltage $V_{GS}$ at $T = 1.3~\mathrm{K}$. Two distinct regimes of high and low p-type transconductance are evident. We attribute the low transconductance regime to modulation of hole density in the bulk of the exfoliated flake, and the high transconductance regime to modulation of a hole accumulation layer. A threshold voltage of $V_{T} = -37~V$ is identified from the onset of high transconductance. There is minimum hysteresis in $G_{xx}$ versus $V_{GS}$, indicating that the chemical potential at the SiO$_{2}$/SnSe interface can be modulated with minimal charge trapping. The field effect mobility $\mu_{FET} = \partial G_{xx} / \partial (C V_{GS})$ and the Hall mobility $\mu_H = R_{yx} / B R_{xx}$ are plotted versus gate voltage $V_{GS}$ in Fig. 3d), where $C = 11.5~\mathrm{nF cm^{-2}}$ is the back gate capacitance and a single carrier type model is assumed here for simplicity. A peak field effect mobility of $250~\mathrm{cm^2 V^{-1} s^{-1}}$ and a peak Hall mobility of $210~\mathrm{cm^2 V^{-1} s^{-1}}$ were observed in the hole accumulation regime. As expected, the field effect mobility drops rapidly in the absence of a hole accumulation layer at the SnSe/SiO$_2$ interface due to ineffective modulation of carrier density within the bulk of the exfoliated SnSe flake. The observed field effect mobility in exfoliated SnSe is an order of magnitude greater than that reported to date \cite{Zhao_NR16, Cho17, Pei_16}. \\

\begin{figure}
    \includegraphics [width=3.5in]{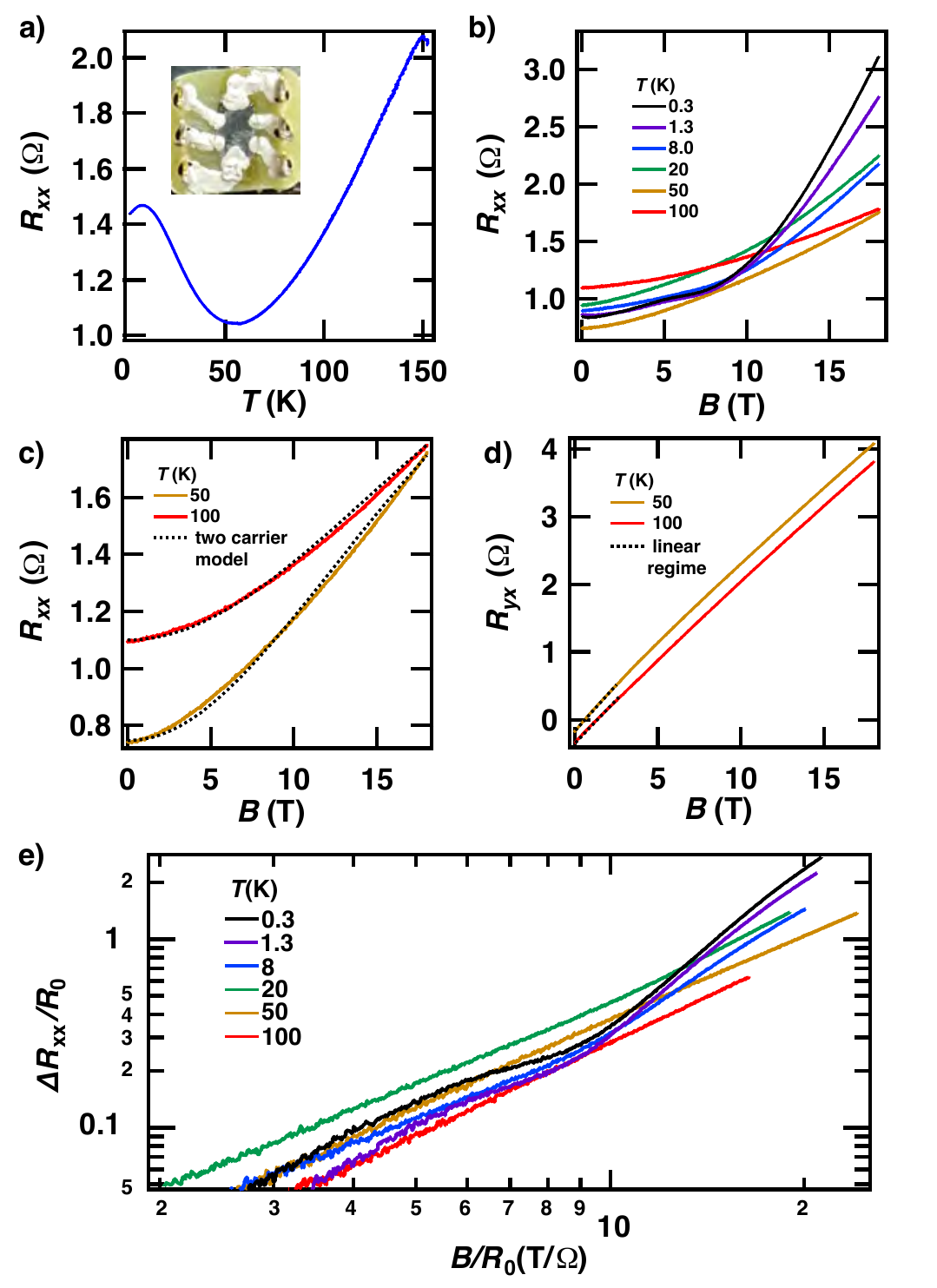}
    \caption{\label{} \textbf{Magnetoresistance.} \textbf{a)} The resistance $R_{xx}$ versus temperature $T$ of a bulk SnSe crystal of thickness $170~\pm12~\mathrm{\mu m}$. \textbf{b)} The resistance $R_{xx}$ versus magnetic field $B$ at different temperatures $T$, showing a clear positive magnetoresistance that increases at low temperature. \textbf{c)} Excellent agreement with a two-carrier model is found for $R_{xx}$ at $T = 50~\mathrm{K}$ and $T= 100~\mathrm{K}$. \textbf{d)} Kohler plot of $\Delta R_{xx}/R_0 =(R_{xx}(B)-R_0)/R_0$ versus $B/R_{0}$. The measured data does not fall on a single universal curve, explicitly showing a violation of Kohler's rule. }
\end{figure}

\textbf{Magnetoresistance}\\

The magnetoresistance of both bulk and exfoliated SnSe was investigated to elucidate the role of the doubly degenerate valence valleys 1) and 2) identified by ARPES.  A bulk SnSe Hall bar, of thickness $170~\pm12~\mathrm{\mu m}$ is shown in Fig. 4a) along with its longitudinal resistance $R_{xx}$ versus temperature $T$. We observed variation in resistance with successive thermal cycles, and measured the magnetoresistance and Hall resistance on a single cool-down. The measured resistance $R_{xx}$ versus magnetic field $B$ at different temperatures is shown in Fig. 4b). Magnetoresistance exceeding 200\% was observed at $B=18~\mathrm{T}$. To analyze the magnetoresistance, we adopt a semi-classical two-carrier model with resistance tensor components
\begin{eqnarray}
R_{xx}  & = &  R_0 \cdot \frac{1+\left( \dfrac{\alpha \mu_1 + \mu_2}{\mu_1 + \alpha \mu_2} \right) \mu_1 \mu_2 B^2}{1+\left( \dfrac{1+\alpha}{\mu_1 + \alpha \mu_2}\right )^2 \mu_1^2 \mu_2^2 B^2} \\
R_{yx} & = & \frac{B}{tpe} \cdot \frac{\left( \dfrac{\mu_1^2+\alpha \mu_2^2}{1+\alpha} \right)+\mu_1^2 \mu_2^2 B^2}{\left( \dfrac{\mu_1 + \alpha \mu_2}{1+\alpha}\right )^2 +\mu_1^2 \mu_2^2 B^2}
\end{eqnarray}
where we assume equal Hall mobility and drift mobility, $\mu_1$, $\mu_2$ and $p_1$, $p_2$ are the mobilities and hole densities in valleys 1 and 2, respectively, $t$ is the sample thickness, $\alpha = p_2/p_1$, $p=p_1+p_2$, and $R_0 = ( e \mu_1 t p_1 + e \mu_2 t p_2 )^{-1}$. Unlike the case of a single-carrier model, the two-carrier model predicts a longitudinal resistance $R_{xx}$ that depends on $B$, and a curvature to the otherwise linear Hall resistance $R_{yx}$ versus $B$.

The two-carrier model is found to give excellent agreement with the measured $R_{xx}$ at $T =50~\mathrm{K}$ and $100~\mathrm{K}$ as shown in Fig. 4c), with lower temperatures exhibiting behaviour that deviates from the semi-classical model. The corresponding measured Hall resistance $R_{yx}$ at $T =50~\mathrm{K}$ and $100~\mathrm{K}$ are shown in Fig. 4d). The non-zero $R_{yx}$ at zero magnetic field is an indication of longitudinal voltage pick-up with the transverse electrodes in the Hall bar. Bipolar magnetic field sweeps (see Supporting Information \cite{supp}) show that the measured $R_{xx}$ has a negligible Hall voltage component over the full-field range, while the Hall resistance $R_{yx}$ is linear over the range $-3~\mathrm{T} < B < +3~\mathrm{T}$ beyond which longitudinal voltage pick-up obscures the curvature of $R_{yx}$ versus $B$ at high field. Thus, a best fit of unipolar $R_{xx}$ versus $0~\mathrm{T}<B<18~\mathrm{T}$ together with the low-field ($B < 3~\mathrm{T}$) Hall resistance slope $\partial R_{yx}/\partial B$ determined by linear fit allows us to estimate the hole valley parameters through equations (1) and (2): $\mu_1 = 2200~\mathrm{cm^{2}V^{-1}s^{-1}}$, $p_1 = 1.3\times10^{17}~\mathrm{cm}^{-3}$, and $\mu_2 = 235~\mathrm{cm^{2}V^{-1}s^{-1}}$, $p_2 = 1.1\times10^{17}~\mathrm{cm}^{-3}$ at $T = 100~\mathrm{K}$; and $\mu_1 = 3400~\mathrm{cm^{2}V^{-1}s^{-1}}$, $p_1 = 1.3\times10^{17}~\mathrm{cm}^{-3}$, and $\mu_2 = 220~\mathrm{cm^{2}V^{-1}s^{-1}}$, $p_2 = 1.6\times10^{17}~\mathrm{cm}^{-3}$ at $T = 50~\mathrm{K}$.

Considering ARPES and magnetoresistance in tandem, we can assign carrier densities $p_1$ and $p_2$ with valleys 1) and 2). The ARPES measurements of Fig.1 and the Supporting Information \cite{supp} show that valley 1) has a larger bandwidth in $K_a$ in comparison with valley 2). In fact, the negligible experimental bandwidth of valley 2) implies it is best described as a quasi-2D pocket. Similarly, the heavier valley 2) can be assigned to the high carrier concentration $p_2$ and low mobility $\mu_2$. The hole densities $p_1$ and $p_2$ imply a chemical potential in close vicinity (15-30~meV at T = 77~K) to the VBM within the bulk of the measured SnSe crystal. However, ARPES spectra reveal a chemical potential 300~meV above VBM at the SnSe surface, indicative of a surface depletion layer. A 300~meV band bending at the observed unintentional doping level of the SnSe gives rise to a $\approx20$~nm surface depletion layer, beyond the reach of surface sensitive methods such as ARPES. 

The presence of two carrier types with different transport properties can be explicitly seen in the Kohler plot of $\Delta R_{xx}/R_0 = (R_{xx}(B)-R_0)/R_0$ versus $B / R_0$ of Fig. 4d). Kohler's rule\cite{kohler} states that $\Delta R_{xx}/R_0$ is a function of the dimensionless parameter $B \tau \propto B/R_0$ alone for semi-classical transport in a single band with a single transport scattering time $\tau$, and thus a universal curve should be obtained versus temperature. The observed deviation from Kohler's rule is further evidence that charge transport cannot be described by a single carrier type with a single scattering time $\tau$.

The order of magnitude difference in hole mobilities $\mu_1$ and $\mu_2$ found by fitting $R_{xx}$ and $\partial R_{yx}/\partial B$ to a two-carrier model implies the transport scattering time $\tau$ differs greatly between valleys 1) and 2). We speculate that the quasi-2D dispersion of valley 2) leads to a lower hole mobility than that in valley 1) because the joint density of states for back-scattering is enhanced by the quasi-2D cylindrical shape of the valley 2). \\

\textbf{Discussion}\\

\begin{figure}
    \includegraphics [width=3.2in]{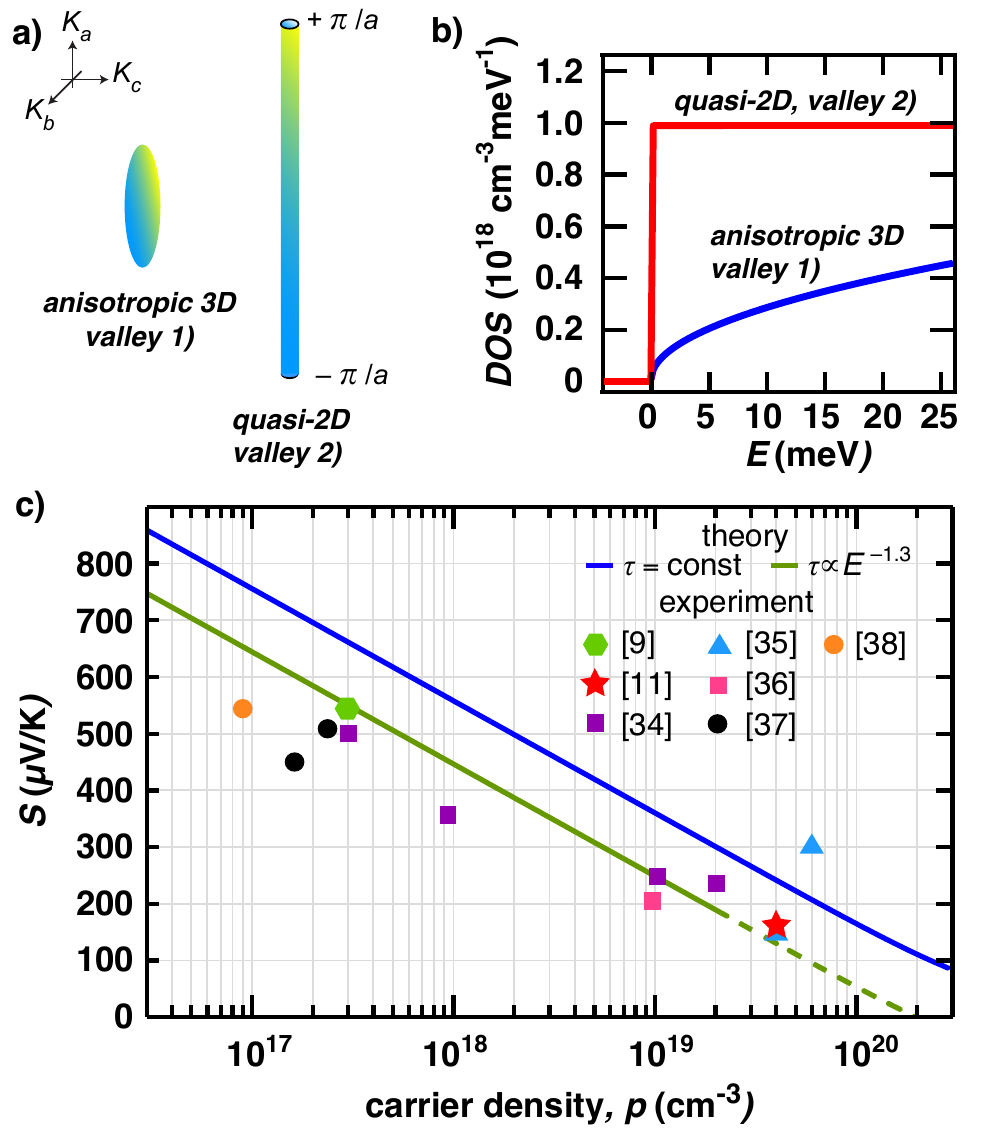}
    \caption{\label{} \textbf{Dimensionality, density of states and Seebeck coefficient.} \textbf{a)} The iso-energy surfaces of valley 1), with anisotropic 3D dispersion, and valley 2), with a quasi-2D dispersion that is flat in out-of-plane.  \textbf{b)} The density of states $\partial p/ \partial E$ versus energy $E$ for the anisotropic 3D dispersion of valley 1) with $m_x = 0.80m_0$, $m_y = 0.25m_0$, and $m_z=0.22m_0$; and the quasi-2D dispersion of valley 2) with $m_y = 0.17$ and $m_z = 0.17$. The well known enhancement of density of states with reduced dimensionality is apparent. \textbf{c)} The Seebeck coefficient $S$ versus total hole density $p$ at room temperature calculated with an energy independent scattering approximation and a scattering time $\tau \propto E^{-1.3}$. Comparison is made against experimental measurements of Seebeck coefficient \cite{Zhao14,Zhao16,Leng,Ge, Chen,Inoue,Serrano}.  }
\end{figure}

We turn our attention to the implication of our findings to understanding the large Seebeck coefficient $S = 160~\mathrm{\mu V K^{-1}}$ at high carrier density $p = 4\times10^{19}~\mathrm{cm^{-3}}$ at room temperature \cite{Zhao16}. The Fermi surface of hole doped SnSe is characterized by a needle-like ellipsoid in valley 1) and a quasi-2D cylinder in valley 2). Notably, a quasi-2D electron dispersion was identified by Hicks and Dresselhaus\cite{hicks} as a means to increase Seebeck coefficient and the $ZT$ figure of merit, with the original proposal calling for a dense stack of quantum wells. Subsequent studies have considered the role of dimensionality in thermoelectricity in more general terms \cite{kim09}. Strong band anisotropy and valley degeneracy has been previously identified as an important factor in the thermoelectric properties of FeAs$_2$ \cite{usui}, LiRh$_2$O$_4$ \cite{arita}, and AgBiSe$_2$ \cite{parker}.

In simple terms, the entropy per charge carrier increases with increasing Fermi surface area at fixed carrier density, as seen in Fig. 5 a) for the anisotropic 3D valley 1) and the quasi-2D valley 2). For example, in the limit of high carrier density leading to degenerate carrier statistics, the Seebeck coefficient from a Sommerfeld approximation for a parabolic 3D valley takes the form $S= (\pi^2/6)\cdot(k_B/e)\cdot( k_B T/E_F )$. Thus, a large density of states favours a large Seebeck coefficient because it leads to a reduction in Fermi level for a given charge carrier density. In the case of SnSe, valley degeneracy, the large out-of-plane mass of valley 1), and the quasi-2D nature of valley 2) all contribute to the enhancement of density of states as shown in Fig. 5 b) using ARPES band masses. The anisotropic 3D valley 1) has a density of states $\partial p / \partial E$, 
\begin{equation}
\left( \frac{ \partial p }{ \partial E } \right)_{3D} = \frac{g }{ 4 \pi^2 }  \cdot \frac{ 2^{3/2} (m_x m_y m_z)^{1/2} }{ \hbar^3 } E^{1/2}
\end{equation}
where $g$ is the spin and valley degeneracy, while the quasi-2D valley 2) has a density of states, 
\begin{equation}
\left( \frac{ \partial p }{ \partial E } \right)_{2D} = \frac{ g }{ 4 \pi a }  \cdot  \frac{ 2 (m_y m_z)^{1/2} } {\hbar^2} 
\end{equation}
where $a$ is the out-of-plane lattice constant. 

We estimate the Seebeck coefficient $S$ for SnSe versus total hole density $p$ at room temperature, first using the energy independent scattering time approximation \cite{hicks} and the valley bandstructure and transport parameters determined in our work. The Seebeck coefficients of the 3D valley 1) and the quasi-2D valley 2) are given by $S_1 = (k_B/e)\cdot ( 5 F_{3/2} / 3 F_{1/2} - \zeta )$ and $S_2 = (k_B/e)\cdot ( 2 F_{1}/F_{0} - \zeta )$, respectively, where $\zeta = E_F / k_BT$ is the reduced chemical potential relative to valence band edge and $F_i = \int_0^{-\infty} ( 1+ \exp(x-\zeta) )^{-1} x^i dx$ is the Fermi-Dirac integral of order $i$ defined for holes in a valence band. The carrier densities $p_1 = ( g / 2^{1/2} \pi^2 ) (m_x m_y m_z)^{1/2} (k_B T)^{3/2} F_{1/2} / \hbar^3 $ and $p_2 = (g/2\pi a) (m_y m_z)^{1/2} k_B T F_{0}/ \hbar^2$ in valleys 1) and 2), respectively, using the band masses determined from ARPES. As Fermi level is varied, the total carrier density $p = p_1 + p_2$ and the total Seebeck coefficient $S = ( S_1 \mu_1 p_1 + S_2 \mu_2 p_2 ) / ( \mu_1 p_1 + \mu_2 p_2 )$, were calculated. We take $\mu_1 / \mu_2 = 9.4$ as inferred from magnetoresistance at $T = 100~\mathrm{K}$, the highest temperature for which we have determined a mobility ratio.

Our simple model estimate for $S$ is compared against experimental measurements of Seebeck coefficient at room temperature in Fig. 5c), including measurements of Seebeck coefficient in polycrystalline p-type SnSe \cite{Ge, Inoue, Serrano}, low-doped single crystal p-type SnSe \cite{Zhao14, Leng} and heavily doped single crystal p-type SnSe \cite{Zhao16, Leng, Chen}. Notably, we find that the Fermi level $E_F$ reaches the valence band edge of SnSe at a carrier density of $p = 5\times10^{19}\mathrm{cm}^{-3}$, owing to the large valence band density of states, from which it follows that degenerate carrier statistics are not reached at room temperature for experimentally reported carrier statistics. The independent scattering time approximation over-estimates the Seebeck coefficient as compared with experimental data over three orders of magnitude of carrier concentration up to the highly doped regime. Improved agreement between experiment and model calculation is reached by inclusion of an energy dependent scattering time $\tau \propto E^{r}$ with a corresponding contribution $\Delta S = r k_B/e$ to the Seebeck coefficient \cite{cai}. In the highest quality SnSe crystals reported to date \cite{maier,yu81}, it was observed that mobility $\mu \propto \tau \propto T^{-2.3}$. With an acoustic phonon population $\propto T$, the scattering rate dependence on charge carrier energy $E$ is estimated to be $1/\tau \propto T\cdot E^{1.3}$ and thus $r=-1.3$ and $\Delta S = -112 ~\mathrm{\mu V / K}$. Further improvements to understanding the thermoelectric performance of SnSe requires a determination of the energy dependence of charge carrier scattering in the valence band valleys of SnSe at experimentally relevant doping levels.

In conclusion, through a detailed experimental study of weakly doped SnSe, we find that the origin of the high Seebeck coefficient of p-type SnSe can be explained through a combination of valence band degeneracy, with one valley being strongly anisotropic and another valley being a quasi-2D band, realizing a layered Hicks-Dresselhaus thermoelectric. Moreover, our experiments show a remarkable difference between black phosphorus, where effective mass is lighter out-of-plane than in the zigzag in-plane direction \cite{Morita}, and the polar analogue SnSe where the out-of-plane mass is significantly heavier than in-plane mass. Our findings motivate further experimental study of the physical properties of heavily doped layered semiconductors for both electronic and thermoelectric applications.\\

\textbf{Acknowledgements}\\

V. T., I. F., N. H., M.P., G. G. and T. S acknowledge NSERC, CIFAR, Hydro-Quebec, L'Institut d'Energie Trottier and the Canada Research Chairs program for financial support. A portion of this work was performed at the National High Magnetic Field Laboratory which is supported by National Science Foundation Cooperative Agreement No. DMR-0084173, the State of Florida, and the Department of Energy. We thank G. Jones, J.-H. Park and T. P. Murphy for outstanding technical support at the National High Magnetic Field Laboratory.  B. S., N. E., A. F. and A. G. acknowledge European Research Council grant no. 648589 `SUPER-2D', funding from Deutsche Forschungsgemeinschaft projects CRC 1238 (project A1) and GR 3708/2-1. D.R. and A.G. acknowledge the G-RISC program. We thank SOLEIL, HZB BESSY and Elettra for the allocation of synchrotron radiation beam times. The stay at the Elettra synchrotron has been supported by the CERIC-ERIC consortium. The Synchrotron SOLEIL is supported by the Centre National de la Recherche Scientifique (CNRS) and the Commissariat \`a l'\'Energie Atomique et aux \'Energies Alternatives (CEA), France.\\

\textbf{Appendix A: Crystal Growth}\\

SnSe crystals were grown from a melt of 6N purity Sn and Se in stoichiometric proportion by the Bridgman method in double-walled quartz ampoules by pulling from a high temperature zone ($\sim$~1180~K) at a rate of 0.56 mm/hr through a linear temperature gradient of 30~K/cm into a colder zone, followed by slow cooling \cite{lada}.\\

\textbf{Appendix B: Density Functional Theory}\\

The electronic band structure of SnSe has been calculated within the pseudopotential plane wave density functional theory (DFT) approach with GW corrections to reproduce the band gap. All computations were performed with use of the ABINIT package~\cite{Gonzeetal.2016}. The atomic coordinates and lattice parameters have been fixed at their experimental values~\cite{Wiedemeier78}. The DFT results were obtained with the Perdew-Burke-Ernzerhof~\cite{Perdew1996} exchange-correlation functional, Troullier-Martins~\cite{NTroullier1991} norm-conserving pseudopotentials, a plane wave set with a cutoff energy of 20 Hartree to represent the wave function and a 6x6x4 Monkhorst-Pack k-point grid for Brillouin zone sampling. The resulting Kohn-Sham data was used as a starting point for the quasiparticle computations. For the G$_0$W$_0$ calculations we used a plane wave set with a cutoff energy of 20 Hartree for the wave function, 6 Hartree to calculate the screening and 80 Hartree for the exchange part of the self-energy. 692 bands has been used for both screening and self-energy calculations and the Brillouin zone was sampled with a 6x6x4 k-point grid. The plasmon-pole model was used for the integration over frequency in the self-energy. To study the position and shape of the multiple energy pockets in k-space, the valence band surface has been calculated on a dense 100x100 k-point grid within the DFT (Fig. 1c).\\

\textbf{Appendix C: ARPES}\\

ARPES measurements were carried out at the ANTARES beamline~\cite{ANTARES14} of the SOLEIL synchrotron and at the UE112 beamline of BESSY on SnSe single crystals with linear polarized light. The crystal surfaces were prepared in-situ by top post cleavage in UHV. Immediately after the cleave we determined the high-symmetry directions through low-energy electron diffraction and rough ARPES maps. The temperatures of the measurements were equal to 35~K for the ARPES maps in the cleavage plane (measured at UE112) and equal to 100K for the out-of-plane dispersion measuements (measured at ANTARES and UE112). The ARPES spectra were obtained using a hemispherical Scienta R4000 analyser. The total energy resolution was $\sim$12 meV.\\

\textbf{Appendix D: Optical Reflectance Spectroscopy}\\

Optical reflectance spectra were acquired at the SISSI endstation~\cite{SISSI07} of the ELETTRA synchrotron with unpolarized light on freshly cleaved SnSe surfaces over a spectral range of $50-17000~\mathrm{cm^{-1}}$ with a Bruker Vertex 70v spectrometer equipped with a variety of beamsplitters and detectors. The measurements were carried out at room temperature. The optical conductivity is extracted from the reflectivity data with the help of Kramers-Kronig transformations, with standard extrapolation procedures \cite{dressel}. \\

\textbf{Appendix E: Device Fabrication}\\

SnSe was exfoliated on to a pre-patterned oxidized silicon wafer ( $5~\mathrm{m \Omega-cm}$ As-doped Si with 300~nm dry SiO$_2$ ) with an adhesive tape method in a glove box with $<1$~ppm H$_2$O and O$_2$ concentration to prevent oxidation \cite{zhaoreview}. The oxidized silicon was treated with hexamethyldisilazane (HMDS) prior to exfoliation, to suppress charge transfer doping. Exfoliated flakes were contacted using a standard electron beam lithography, metal evaporation and lift-off technique. Electrodes were arranged in a two-point or van der Pauw geometry. Once fabricated, the SnSe was encapsulated in a glove box environment by spin-coating 300~nm of copolymer (methyl methacrylate) and 200~nm of polymer (polymethyl methacrylate) followed by an annealing step at 170~C for 15~min. Bulk SnSe samples were cleaved in ambient conditions and contacted by a combination of metal evaporation through a shadow mask and manual application of Ag paste.\\

\textbf{Appendix F: Transport Measurements}\\

Quasi-dc measurements were performed at room temperature in a vacuum probe station with a semiconductor parameter analyzer. Cryogenic measurements were performed using standard lock-in techniques in a He$^4$ variable temperature insert fitted with an 8~T superconducting solenoid. Exfoliated samples were illuminated with red light during cool down with red LEDs biased at 1~mA, and mounted in close proximity to the exfoliated flakes. In all field effect measurements, gate leakage currents were monitored and did not exceed 1~nA. Exfoliated samples were characterized with ac source-drain currents of 1~$\mu$A, and bulk samples were characterized with ac source-drain currents of 100~$\mu$A.\\

High magnetic field measurements were performed at National High Magnetic Field Laboratory in Tallahassee. An 18~T superconducting solenoid fitted with a He$^3$ cryostat was used for bulk sample measurements. Standard lock-in measurement techniques were used with ac source-drain currents of 10~$\mu$A.\\

\clearpage

\section{Supporting Information}

\subsection{Effective masses from density-functional theory}
\begin{figure*}[htb!]
\includegraphics[width=0.9\textwidth]{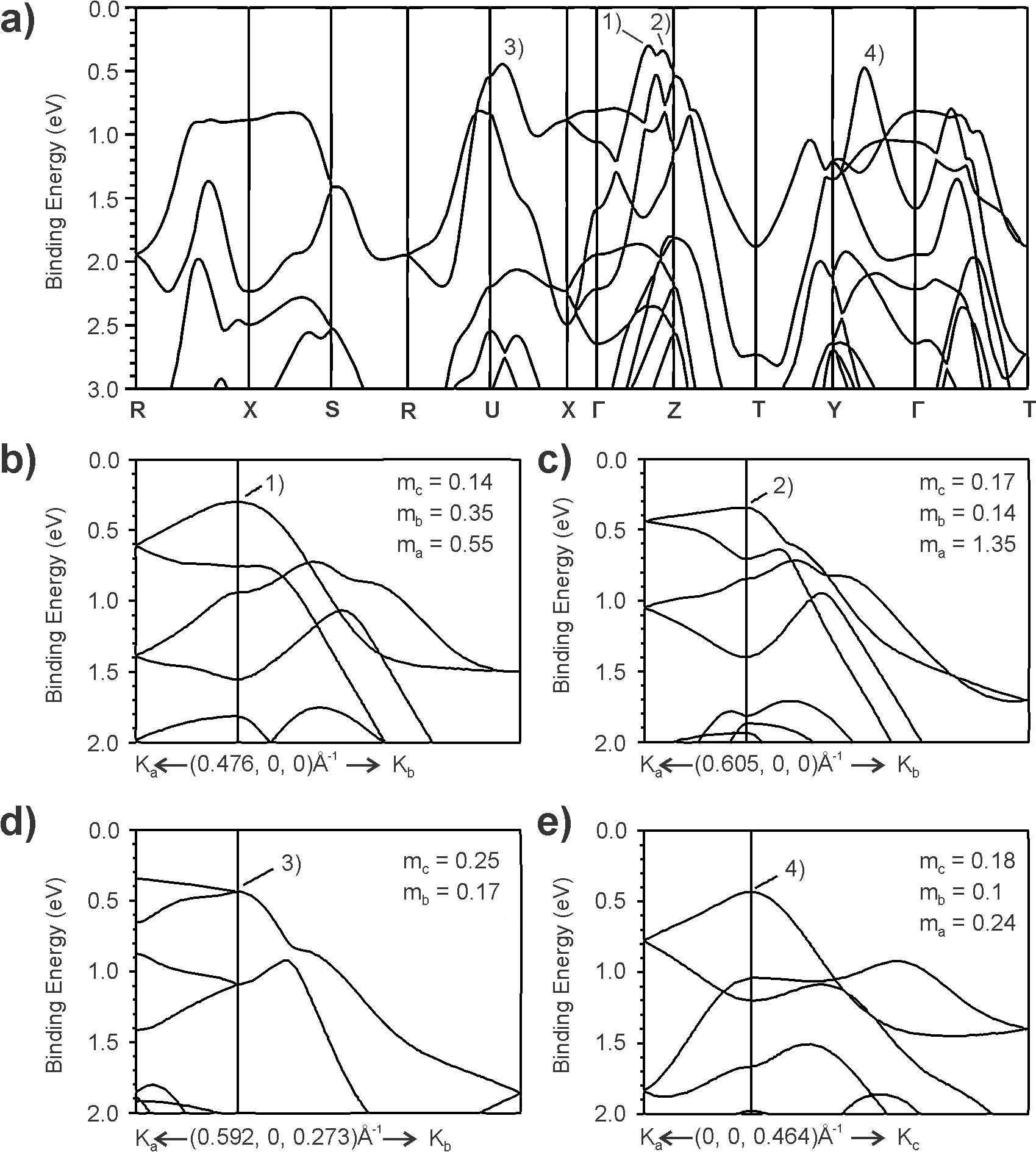}
\caption{(a) Overview of the band structure by DFT. (b)-(e) depict the out-of-plane ($K_a$) and one in-plan ($K_b$) mass for pockets 1)-4).}
\label{fig:masses}
\end{figure*}

Fig.~\ref{fig:masses} depicts an overview of the band structure and the relevant pockets. We performed parabolic fits of the effective mass tensor components for each pocket. As a result we obtain the following masses (all in units of the free electron mass):  valley 1) $m_x$ = 0.55, $m_y$ = 0.35, $m_z$ = 0.14, valley 2) $m_x$ = 1.35, $m_y$ = 0.14, $m_z$ = 0.17 valley 3) $m_y$ = 0.17, $m_z$ = 0.25 (the $m_x$ of this valley is ill-defined since it has a degeneracy at the valley position in $k$ space which splits in two bands along the $k_x$ direction) and valley 4) $m_x$ = 0.24, $m_y$ = 0.10, $m_z$ = 0.18. It is also seen that the valley 3) originates from the valley 2) and from another band rising from the bottom. 

\subsection{Out-of-plane dispersion relation}
\begin{figure*}[htb!]
\includegraphics[width=0.9\textwidth]{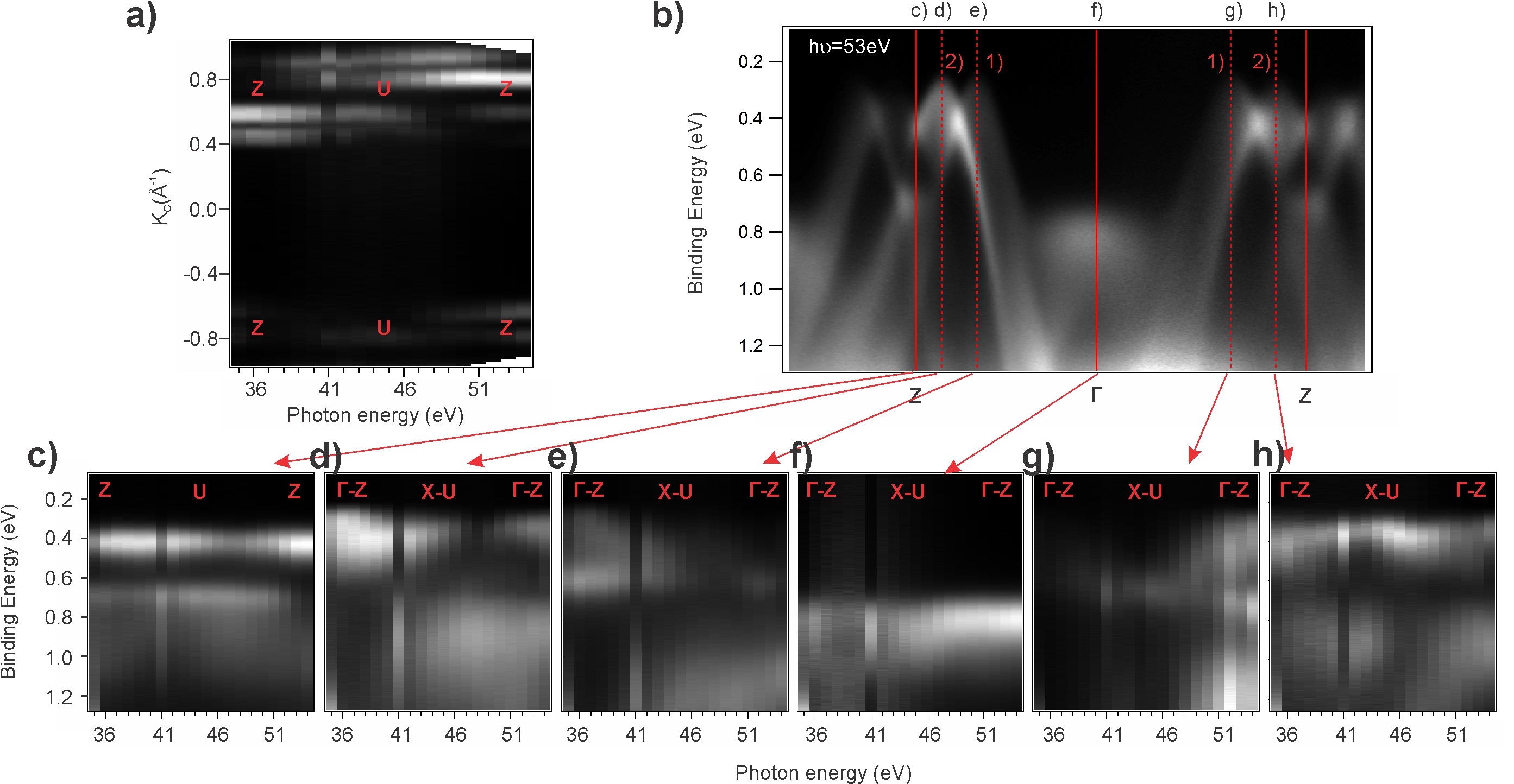}
\caption{a) Constant energy (E=0.3~eV) plot of the momentum scans at different photon energies. b) Spectrum at $h\nu$=53eV indicating the wavevectors where the momentum slices at different photon energies were taken. c-h) Oscillations in the ARPES intensity allow us to assign the $Z$ and $U$ points along the $K_a$ direction. }
\label{fig:arpes}
\end{figure*}

Fig.~\ref{fig:arpes}(a) depicts a constant energy cut thru the top of pockets 1) and 2) at varying photon energy highlighting the matrix element effects. Fig~\ref{fig:arpes}(b) is an ARPES spectrum along $\Gamma Z$ which indicates pockets 1) and 2). Figs.~\ref{fig:arpes}(c-h) shows the photon energy dependence of the pockets 1) and 2). Sweeping the photon energy allows to probe the out-of-plane dispersion relation. 

\subsection{Bipolar magnetotransport}
\begin{figure*}[htb!]
\includegraphics[width=0.9\textwidth]{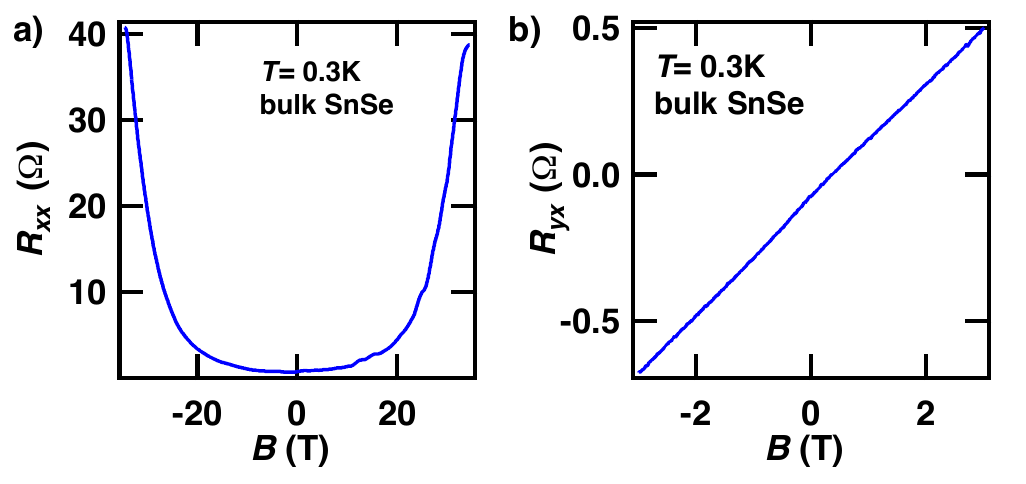}
\caption{a) The longitudinal resistance $R_{xx}$ versus $B$ of a bulk SnSe Hall bar at $T=0.3~\mathrm{K}$. b) The Hall resistance $R_{yx}$ versus $B$ of a bulk SnSe Hall bar sample at $T=0.3~\mathrm{K}$. }
\label{fig:mag}
\end{figure*}

Fig.~\ref{fig:mag}(a) depicts the longitudinal resistance $R_{xx}$ versus $B$ in a bipolar magnetic field sweep of the bulk SnSe Hall bar sample reported in this work at $T=0.3~\mathrm{K}$. The measured $R_{xx}$ is symmetric in $B$, indicating minimum Hall voltage pick-up in the Hall bar. Fig~\ref{fig:mag}(b) depicts the Hall resistance $R_{yx}$ versus $B$ in a bipolar magnetic field sweep of a bulk SnSe Hall bar sample at $T=0.3~\mathrm{K}$. The measured $R_{yx}$ is linear in $B$, over the field range shown, indicating a longitudinal voltage pick-up that produces a vertical offset in $R_{yx}$ versus $B$ but induces a negligible curvature versus $B$ over the field range $-3~\mathrm{T} < B < +3~\mathrm{T}$.

\begin{thebibliography}{10}

\bibitem{PALee}
P. A. Lee \textit{ed.}, Optical and Electrical Properties, Physics and Chemistry of Materials with Layered Structures, vol. \textbf{4} (D. Riedel Pub. Co., Dordrecht, 1976).

\bibitem{VGrasso}
V. Grasso \textit{ed.}, Electronic Structure and Electronic Transitions in Layered Materials, Physics and Chemistry of Materials with Low-Dimensional Structures, series A, (D. Riedel Pub. Co., Dordrecht, 1986).


\bibitem{Wiedemeier78}
H. Wiedemeier, H. G. v. Schnering, Refinement of the structures of GeS, GeSe, SnS and SnSe, Z. Kristallogr. \textbf{148}, 295 (1978).

\bibitem{Morita}
A. Morita,  Semiconducting black phosphorus. Appl. Phys. A \textbf{39}, 227-242 (1986).

\bibitem{Li_NN2014}
L. Li, et al, Black phosphorus field effect transistors, Nat. Nanotechnol. \textbf{9}, 372 (2014).

\bibitem{Liu_ACN2014}
H. Liu, et al, Phosphorene: an unexplored 2D semiconductor with a high hole mobility, ACS Nano \textbf{8}, 4033 (2014).

\bibitem{Xia_NC2014}
F. Xia,  H. Wang, and Y. Jia, Rediscovering black phosphorus as an anisotropic layered material for optoelectronics and electronics, Nat. Commun. \textbf{5}, 4458 (2014).

\bibitem{Gomez}
A. Castellanos-Gomez, Black Phosphorus: Narrow Gap, Wide Applications, J. Phys. Chem. Lett. \textbf{6}, 4280 (2015).



\bibitem{Zhao14}
L.-D. Zhao, S.-H. Lo, Y. Zhang, H. Sun, G. Tan, C. Uher, G.J. Snyder, C. Wolverton, V. P. Dravid, M. G. Kanatzidis, Ultralow thermal conductivity and high thermoelectric figure of merit in SnSe crystals, Nature \textbf{508}, 373 (2014).

\bibitem{Li2015}
C.W. Li, J. Hong, A. F. May, D. Bansal, S. Chi, T. Hong, G. Ehlers and O. Delaire, Orbitally driven giant phonon anharmonicity in SnSe, Nature Phys. \textbf{11}, 1063 (2015).

\bibitem{Zhao16}
L.-D. Zhao, G. Tan,  S. Hao, J. He, Y. Pei, H. Chi, H. Wang, S. Gong, H. Xu, V. P. Dravid, C. Uher, G. J. Snyder, C. Wolverton, and M. G. Kanatzidis, Ultrahigh power factor and thermoelectric performance in hole-doped single-crystal SnSe, Science \textbf{351}, 141 (2016).

\bibitem{Duong} A. T. Duong, V. Q. Nguyen, G. Duvjir, V. T. Duong, S. Kwon, J. Y. Song, J. K. Lee, J. E. Lee, S.-D. Park, T. Min, J. Lee, J. Kim and S. Cho, Achieving ZT=2.2 with Bi-doped n-type SnSe single crystals, Nat. Comm. \textbf{7}, 13713 (2016).

\bibitem{Snyder} G. J. Snyder and E. S. Toberer, Complex thermoelectric materials, Nature Mat. \textbf{7}, 105 (2008).

\bibitem{Cuong_15} D. D. Cuong, S. H. Rhim, J.-H. Lee and S. C. Hong, Strain effect on electronic structure and thermoelectric properties of orthorhombic SnSe: a first principles study, AIP Advances \textbf{5}, 117147 (2015).

\bibitem{hicks}
L. D. Hicks and M. S. Dresselhaus, Effect of quantum-well structures on the thermoelectric figure of merit, Phys. Rev. B \textbf{47}, 12727 (1993).

\bibitem{maier}
H. Maier and D. R. Daniel, SnSe single crystals: sublimation growth, deviation from stoichiometry and electrical properties, J. Elec. Mat. \textbf{6}, 693 (1977).

\bibitem{yu81}
J. G. Yu, A. S. Yue and O. M. Stafsudd, Growth and electronic properties of the SnSe semiconductor, J. Cryst. Growth \textbf{54}, 248 (1981).

\bibitem{Bhatt_89}
V. P. Bhatt, K. Gireesan and G. R. Pandya, Growth and characterization of SnSe and SnSe$_2$ crystals, J. Cryst. Growth \textbf{96}, 649 (1989).

\bibitem{Asanabe}
S. Asanabe, Electrical Properties of Stannous Selenide, J. Phys. Soc. Jap. \textbf{14}, 281 (1959).

\bibitem{Mitzi}
D.B. Mitzi, L. L. Kosbar, C. E. Murray, M. Kopel and A. Afzali, High-mobility ultrathin semiconducting films prepared by spin coating, Nature \textbf{428}, 299 (2004).

\bibitem{Zhao_NR16}
S. Zhao, H. Wang, Y. Zhou, L. Liao, Y. Jiang, X. Yang, G. Chen, M. Lin, Y. Wang, H. Peng and Z. Liu, Controlled synthesis of single-crystal SnSe nanoplates, Nano Res. \textbf{8}, 288 (2015).

\bibitem{Cho17}
S.-H. Cho, K. Cho, N.-W. Park, S. Park, J.-H. Koh and S.-K. Lee, Multi-Layer SnSe Nanoflake Field effect Transistors with Low-Resistance Au Ohmic Contacts, Nanoscale Res. Lett. \textbf{12}, 273 (2017).

\bibitem{Pei_16}
T. Pei, L. Bao, R. Ma, S. Song, B. Ge, L. Wu, Z. Zhou, G. Wang, H. Yang, J. Li, C. Gu, C. Shen, S. Du, and H.-J. Gao, Epitaxy of ultrathin SnSe single crystals on polydimethylsiloxane: in-plane electrical anisotropy and gate-tunable thermopower, Adv. Electron. Mater. \textbf{2}, 1600292 (2016).


\bibitem{Wang17}
C. W. Wang, Y. Y. Y. Xia, Z. Tian, J. Jiang, B. H. Li, S. T. Cui, H. F. Yang, A. J. Liang, X. Y. Zhan, G. H. Hong, S. Liu, C. Chen, M. X. Wang, L. X. Yang, Z. Liu, Q. X. Mi, G. Li, J. M. Xue, Z. K. Liu, and Y. L. Chen, Photoemission study of the electronic structure of valence band convergent SnSe, Phys. Rev. B. \textbf{96}, 165118 (2017).

\bibitem{Lu17}
Q. Lu, M. Wu, D. Wu, C. Chang, Y.-P. Guo, C.-S. Zhou, W. Li, X.-M. Ma, G. Wang, L.-D. Zhao, L. Huang, C. Liu and J. He, Unexpected Large Hole Effective Masses in SnSe Revealed by Angle-Resolved Photoemission Spectroscopy, Phys. Rev. Lett. \textbf{119}, 116401 (2017).

\bibitem{supp}
See Supplemental Material at [URL will be inserted by publisher] for a summary of effective mass determinations from DFT bandstructure calculations, analysis of out-of-plane dispersion relations from ARPES, and bipolar magnetotransport measurements.




\bibitem{cardona77-snse}
H. R. Chandrasekhar, R. G. Humphreys, U. Zwick, and M. Cardona, Infrared and Raman spectra of the IV-VI compounds SnS and SnSe, Phys. Rev. B, \textbf{15}, 2177, (1977).

\bibitem{aamir17-phonons}
A. Shafique and Y.-H. Shin, Thermoelectric and phonon transport properties of two-dimensional IV-VI compounds, Sci. Rep. \textbf{7}, 506 (2017).

\bibitem{elkorashy86-snse}
A. M. Elkorashy, Optical absorption in tin monoselenide single crystal, J. Phys. Chem. Sol. \textbf{47}, 497 (1986).

\bibitem{martin90-snse}
M. Parenteau and C. Carlone, Influence of temperature and pressure on the electronic transitions in SnS and SnSe semiconductors, Phys. Rev. B, \textbf{41}, 5227 (1990).

\bibitem{CRC}
CRC Handbook of Chemistry and Physics.

\bibitem{Lang}
D. V. Lang, R. A. Logan, and M. Jaros, Trapping characteristics and a donor-complex (DX) model for the persistent-photoconductivity
trapping center in Te-doped Al$_x$Ga$_{1-x}$As, Phys. Rev. B \textbf{19}, 1015 (1979).


\bibitem{kohler}
M. Kohler, Zur magnetischen Widerstands\"anderung reiner Metalle, Ann. der Phys. \textbf{32}, 211 (1938).

\bibitem{Leng}
H.-Q. Leng, M. Zhou, J. Zhao, Y.-M. Han and L.-F. Li, The thermoelectric performance of anisotropic SnSe doped with Na,  RSC Adv. \textbf{6}, 9112 (2016).

\bibitem{Ge}
Z.-H. Ge, D. Song, X. Chong, F. Zheng, L. Jin, X. Qian, L. Zheng, R. E. Dunin-Borkowski, P. Qin, J. Feng, and L.-D. Zhao, Boosting the Thermoelectric Performance of (Na,K)-Codoped Polycrystalline SnSe by Synergistic Tailoring of the Band Structure and Atomic-Scale Defect Phonon Scattering, J. Am. Chem. Soc. \textbf{139}, 9714 (2017).

\bibitem{Chen}
C.-L. Chen, H. Wang, Y.-Y. Chen, T. Daya and G. Jeffrey Snyder, Thermoelectric properties of p-type polycrystalline SnSe doped with Ag, J. Mater. Chem. A \textbf{2}, 11171 (2014).

\bibitem{Inoue}
T. Inoue, H. Hiramatsu, H. Hosono, and T. Kamiyar, Heteroepitaxial growth of SnSe films by pulsed laser deposition using Se-rich targets, Appl. Phys. Lett. \textbf{118}, 205302 (2015).

\bibitem{Serrano}
F. Serrano-S\'anchez, M. Gharsallah, N. M. Nemes, F. J. Mompean, J. L. Mart\'inez, and J. A. Alonso, Record Seebeck coefficient and extremely low thermal conductivity in nanostructured SnSe, Appl. Phys. Lett. \textbf{106}, 083902 (2015).

\bibitem{kim09} 
R. Kim, S. Datta and M. S. Lundstrom, Influence of dimensionality on thermoelectric device performance, J. Appl. Phys. \textbf{105}, 034506 (2009).

\bibitem{usui}
H. Usui, K. Suzuki, K. Kuroki, S. Nakano, K. Kudo, and M. Nohara, Large Seebeck effect in electron-doped FeAs$_2$
driven by a quasi-one-dimensional pudding-mold-type band, Phys. Rev. B \textbf{88}, 075140 (2013).

\bibitem{arita}
R. Arita, K. Kuroki, K. Held, A. V. Lukoyanov, S. Skornyakov, and V. I. Anisimov, Origin of large thermopower
in LiRh$_2$O$_4$: Calculation of the Seebeck coefficient by the combination of local density approximation and
dynamical mean-field theory, Phys. Rev. B \textbf{78}, 115121 (2008).

\bibitem{parker}
D. S. Parker, A. F. May, and D. J. Singh, Benefits of Carrier-Pocket Anisotropy to Thermoelectric Performance:
The Case of p-Type AgBiSe$_2$, Phys. Rev. Appl. \textbf{3}, 064003 (2015).

\bibitem{cai}
J. Cai and G. D. Mahan, Effective Seebeck coefficient for semiconductors, Phys. Rev. B \textbf{74}, 075201 (2006).

\bibitem{lada}
V.I. Shtanov and  L.V. Yashina, On the Bridgman growth of lead-tin selenide crystals with uniform tin distribution, J. Cryst. Growth \textbf{311}, 3257 (2009).

\bibitem{Gonzeetal.2016}
X. Gonze \textit{et al.}, Recent developments in the ABINIT software package, Comp. Phys. Comm. \textbf{205}, 106 (2016).

\bibitem{Perdew1996}
J. P. Perdew, K. Burke and M. Ernzerhof, Generalized Gradient Approximation Made Simple, Phys. Rev. Lett. \textbf{77}, 3865 (1996). 

\bibitem{NTroullier1991}
N. Troullier and J.L. Martins, Efficient pseudopotentials for plane-wave calculations, Phys. Rev. B \textbf{43}, 1993 (1991).

\bibitem{ANTARES14}
J. Avila and M. C. Asensio, First NanoARPES User Facility Available at SOLEIL: An Innovative and Powerful Tool for Studying Advanced Materials, Synchrotron Radiation News {\bf 27}, 24, (2014).

\bibitem{SISSI07}
S. Lupi and A. Nucara and A. Perucchi and P. Calvani and M. Ortolani and L. Quaroni and M. Kiskinova, Performance of SISSI, the infrared beamline of the ELETTRA storage ring, J. Opt. Soc. Am. B {\bf 24}, 959 (2007).

\bibitem{dressel}
M. Dressel and G. Gr\"uner, Electrodynamics of Solids: optical properties of electrons in matter, (Cambridge University Press, Cambridge, 2002).

\bibitem{zhaoreview}
L.-D. Zhao, C. Chang, G. Tan and M. G. Kanatzidis, SnSe: a remarkable new thermoelectric material, Energy Environ. Sci. \textbf{9}, 3044 (2016).

\end{thebibliography}

\end{document}